\documentclass[10pt,conference]{IEEEtran}

\makeatletter
\def\ps@headings{%
\def\@oddhead{\mbox{}\scriptsize\rightmark \hfil \thepage}%
\def\@evenhead{\scriptsize\thepage \hfil \leftmark\mbox{}}%
\def\@oddfoot{}%
\def\@evenfoot{}}
\makeatother
\usepackage{cite}
\usepackage{amsmath,amssymb,amsfonts}
\usepackage{algorithmic}
\usepackage{graphicx}
\usepackage{textcomp}
\usepackage{xcolor}
\usepackage{balance}
\usepackage{url}
\usepackage{newtxtext,newtxmath} 
\usepackage{tikz}

\newcommand\copyrighttext{%
  \footnotesize \textcopyright \the\year{} IEEE. Personal use of this material is permitted. Permission from IEEE must be obtained for all other uses, including reprinting/republishing this material for advertising or promotional purposes, collecting new collected works for resale or redistribution to servers or lists, or reuse of any copyrighted component of this work in other works. The final version of record will be available via IEEE Xplore®.}

\newcommand\copyrightnotice{%
\begin{tikzpicture}[remember picture,overlay]
\node[anchor=south,yshift=12pt] at (current page.south) {\fbox{\parbox{\dimexpr0.75\textwidth-\fboxsep-\fboxrule\relax}{\copyrighttext}}};
\end{tikzpicture}%
}

\def\BibTeX{{\rm B\kern-.05em{\sc i\kern-.025em b}\kern-.08em
    T\kern-.1667em\lower.7ex\hbox{E}\kern-.125emX}}
    
\begin{document}

\title{ICS-SimLab: A Containerized Approach for Simulating Industrial Control Systems for Cyber Security Research\\
}

\author{
  \IEEEauthorblockN{Jaxson Brown$\dagger$, Duc-Son Pham$\dagger$, Sie-Teng Soh$\dagger$, Foad Motalebi$\ddagger$, Sivaraman Eswaran$\ddagger$, and Mahathir Almashor$\S$}
  \IEEEauthorblockA{
    \textit{$\dagger$ School of Electrical Engineering, Computing and Mathematical Sciences, Curtin University, Western Australia}\\
    \textit{$\ddagger$ Department of Electrical and Computer Engineering, Curtin University, Malaysia}\\
    \textit{$\S$ Commonwealth Scientific and Industrial Research Organisation (CSIRO) Energy, Perth, Western Australia }\\
    jokesene@outlook.com, 
    dspham@ieee.org,
    s.soh@curtin.edu.au, \\
    foad.m@curtin.edu.my,
    sivaraman.eswaran@curtin.edu.my,
    mahathir.almashor@csiro.au
  }
}

\maketitle

\copyrightnotice
\begin{abstract}
Industrial Control Systems (ICSs) are complex interconnected systems used to manage process control within industrial environments, such as chemical processing plants and water treatment facilities. 
As the modern industrial environment moves towards Internet-facing services, ICSs face an increased risk of attacks that necessitates ICS-specific Intrusion Detection Systems (IDS). The development of such IDS relies significantly on a simulated testbed as it is unrealistic and sometimes hazardous to utilize an operational control system.
Whilst some testbeds have been proposed, they often use a limited selection of virtual ICS simulations to test and verify cyber security solutions. There is a lack of investigation done on developing systems that can efficiently simulate multiple ICS architectures. Currently, the trend within research involves developing security solutions on just one ICS simulation, which can result in bias to its specific architecture.

We present ICS-SimLab, an end-to-end software suite that utilizes Docker containerization technology to create a highly configurable ICS simulation environment. This software framework enables researchers to rapidly build and customize different ICS environments, facilitating the development of security solutions across different systems that adhere to the Purdue Enterprise Reference Architecture. To demonstrate its capability, we present three virtual ICS simulations: a solar panel smart grid, a water bottle filling facility, and a system of intelligent electronic devices. Furthermore, we run cyber-attacks on these simulations and construct a dataset of recorded malicious and benign network traffic to be used for IDS development.
\end{abstract}


\section{Introduction}
Industrial Control Systems (ICSs) are complex, highly interconnected systems used to manage process control and automation across industrial environments. It is a general term that includes systems such as Supervisory Control and Data Acquisition (SCADA) systems, Distributed Control Systems (DCS), as well as components such as Programmable Logic Controllers (PLC) and Remote Terminal Units (RTU) \cite{drias2015analysis}. ICSs are found in facilities such as water treatment plants, transport infrastructure, power generation systems and intelligent electronic devices \cite{ani2021design}. Systems that require real-time data monitoring and control are typically implemented as an ICS.

With the convergence of Information Technology (IT) and Operational Technology (OT), many ICS environments are increasingly exposed to cyber threats. This is especially prevalent with the use of Internet-facing services, which introduces remote attack vectors \cite{hu2018survey}. As such, there is growing demand for innovative and effective cyber security solutions for ICS technology. Furthermore, these solutions must be designed so that they do not impact the availability of these systems. Such a solution includes Intrusion Detection Systems (IDS), which monitor network and/or host activity for malicious behavior. IDS software is an example of technology that doesn't limit the availability of a system, making it suitable for ICS environments.

Historical examples of past ICS cyber attacks highlight the importance of implementing innovative security solutions. A commonly referenced example includes the Stuxnet attack. Stuxnet was a highly sophisticated malware that was designed to specifically target ICS environments \cite{falliere2011w32, chen2011lessons}. It's known to be the first malware constructed to cause physical harm, and has been referred to as a ``cyber warfare weapon'' \cite{langner2011stuxnet}. In 2010, Stuxnet was responsible for an attack on Iranian uranium enrichment facilities, which involved the physical destruction of centrifuges used in the enrichment process.

BlackEnergy is another key malware example that demonstrates the extreme consequences of ICS cyber attacks. In 2015, the third iteration of the BlackEnergy malware was used in a large-scale attack on Ukrainian power grids. This malware abused exploits in outdated Microsoft Office documents to establish remote backdoor access into power grid operations\cite{case2016analysis}. This attack caused power outages across Ukraine and resulted in approximately 225000 citizens without power \cite{alert2016cyber}.

Considering the magnitude of past attacks, it's clear that cyber protection and solution development must be a priority for the modern ICS landscape. To be able to develop and evaluate such security solutions, researchers require testbeds that replicate real-world control systems. However, testing on live ICS infrastructure presents significant challenges due to safety, availability, and confidentiality concerns. As a result, ICS simulation environments are being used as an alternative. These environments replicate ICS components and network behavior to provide realistic testbeds for general ICS cyber security research.

Within this field, researchers often use previously built simulations from other works. This is common practice as it allows researchers to avoid hardware related issues and other technical problems involved with setting up a testbed \cite{dehlaghi2023icssim}. Researchers investigating Machine Learning (ML)-based IDS development can also leverage existing datasets specifically designed for training. These datasets are typically produced by running cyber-attacks on ICS simulations whilst recording network activity. Through using these datasets, researchers can avoid setting up a simulation entirely \cite{mubarak2022industrial, morris2014industrial}. However, there are drawbacks to using existing simulations and datasets. There is a lack of flexibility in defining custom test conditions, as well as custom cyber-attacks \cite{dehlaghi2023icssim}. Malware attacks are often evolving, and previous datasets do not reflect that evolution \cite{teixeira2018scada}. There is also the major issue of bias within developed solutions. For instance, if a developed IDS software works well on one type of ICS architecture, there is a high probability that the same software won't be effective on a different type of ICS.

We present ICS-SimLab, a software suite capable of simulating various types of ICS environments. This software utilizes Docker containerization technology to realize the different components of ICS and SCADA systems, specifically Programmable Logic Controllers (PLCs), Human Machine Interfaces (HMIs), sensors and actuators. We also implement Hardware-in-the-Loop modules to replicate the physical processes involved in these control systems. Due to the efficiency of containers, ICS-SimLab can be run entirely on a single computer with relatively minimal computational requirements. Previous works often use other encapsulation technologies, such as virtual machines. However it has been demonstrated that these technologies have substantial computational requirements and may cause issues if running on a single host \cite{de2022development}.

Furthermore, ICS-SimLab employs a highly configurable system that enables researchers to customize a wide range of ICS architectures. This is achieved through a JSON-based configuration format that defines all key ICS features, including descriptions of devices, communications, and control logic. ICS-SimLab uses these predefined JSON files to construct Docker containers representing the various components of an ICS, which are then deployed within a virtual network to form the simulation. Several external tools are also used to virtualize other aspects of an ICS environment. 

This paper focuses on usability and reproducibility throughout the design and implementation process. Reproducibility, as highlighted in previous literature \cite{alves2018virtualization}, reinforces the need for accessible and user-friendly ICS simulation tools. Specifically, the configuration system of ICS-SimLab has been designed to be both intuitive and transparent, allowing researchers to use the software with minimal training. Furthermore, abstraction techniques have been employed to improve reliability and reduce the overall complexity of the system.

This paper is organised as follows. In Section~\ref{sec:related} we investigate the literature of previous ICS simulation designs and comment on certain design advantages and flaws. Section~\ref{sec:background} explores the structure of ICS, including the different components and protocols used within ICS environments. Section~\ref{sec:sim-method} describes the design of ICS-SimLab and describes three preconfigured simulations built with the framework. Section~\ref{sec:attack-method} highlights a practical application of the simulation software through investigating the effects of different cyber-attacks. In this section we also present several datasets that can be used for IDS developments. Finally, Section \ref{sec:conclusion} discusses areas of improvement for future work and concludes the results of this paper.

\section{Related Work}
\label{sec:related}
In this section we review three categories of ICS testbeds as defined in \cite{conti2021survey}: physical, virtual and hybrid. We examine the strengths of the testbed designs and comment on any limitations. Table \ref{tab:literature} contains a summary of relevant literature.

\subsection{Physical Testbeds}
Physical ICS simulations are built using real ICS devices, such as industry-grade Programmable Logic Controllers (PLCs) and Intelligent Electronic Devices (IEDs). They are often designed to be fully working, miniaturized control systems.

Teixeira et al. \cite{teixeira2018scada} have built an ICS testbed representing a simple water tank facility. Their testbed involves a physical water tank equipped with sensors, actuators and a PLC to control water levels within the tank. The purpose of this testbed is to help emulate “real-world industrial systems as closely as possible without replicating an entire plant or assembly system” \cite{teixeira2018scada}, although the simplicity of the simulation has been criticized in \cite{de2022development}.

\begin{table}[htbp]
    \caption{Brief outline of popular ICS simulations and frameworks within literature.}
        \begin{center}
                \begin{tabular}{|c|c|p{5cm}|}
                \hline
                \textbf{Category} & \textbf{Paper} & \textbf{Description} \\
                \hline
                Physical Testbed & \cite{teixeira2018scada} & Created a physical water tank ICS and tested classical ML for intrusion detection for reconnaissance attacks. \\
                \hline
                Physical Testbed & \cite{mathur2016swat} & A physical testbed that implements a 6 stage water purification process. \\
                \hline
                Physical Testbed & \cite{morris2011control} & A testbed with seven different physical processes. \\
                \hline
                Virtual Testbed & \cite{de2022development} & VM-based nuclear power plant simulation. \\
                \hline
                Virtual Testbed & \cite{boakye2023securing} & Builds a substation tested using Docker containers and Modbus-TCP. \\
                \hline
                Virtual Testbed & \cite{morris2015industrial} & Builds a physical and virtual gas pipeline testbed. \\
                \hline
                Virtual Testbed & \cite{candia2024tinyics} & An open-source ICS simulator that models industrial communication. \\
                \hline
                Hybrid Testbed & \cite{noorizadeh2021cyber} & Creates a hybrid simulation of the Tennessee Eastman (TE) plant, along with ML modules for IDS development. \\
                \hline
                Hybrid Testbed & \cite{mubarak2022industrial} & Hybrid waste water simulation using real PLCs and simulated analogue I/O. \\
                \hline
                Framework & \cite{dehlaghi2023icssim} & Creates an OOP-based framework for simulating ICS in Docker containers. \\
                \hline
                Framework & \cite{alves2018virtualization} & VM-based framework that uses open-source software to virtualize different ICS components. \\
                \hline
                Framework & \cite{formby2018lowering} & A framework for visualizing ICS environments using the Unity game engine. \\
                \hline
            \end{tabular}
        \label{tab:literature}
    \end{center}
\end{table}

Mathur and Tippenhauer \cite{mathur2016swat} introduced a small-scale water purification ICS facility, designed specifically for cyber security research. The testbed, named the SwaT testbed for “Secure Water Treatment”, features a six-stage water purification process. Its sophisticated design provides a highly realistic testbed platform, however it also makes it impractical for independent replication.

Morris et al. \cite{morris2011control} developed a physical testbed consisting of seven miniaturized control systems, which incorporated both serial and Ethernet communications. The systems using serial communication include a water storage tank, a raised water system, a gas pipeline, a conveyor belt setup, and an industrial blower system. Two Ethernet-based systems were also developed: a steel rolling operation setup and a hybrid smart grid transmission control system.

Overall, physical ICS testbeds provide a valuable environment for comprehensive cyber security research. However, to replicate a physical testbed is very time consuming and often impractical for researchers \cite{dehlaghi2023icssim}. Therefore, it’s worth looking into alternative approaches such as virtual or hybrid simulations.

\subsection{Virtual Simulations}
Virtual ICS simulations are built fully on virtualization technology, such as virtual machines \cite{alves2018virtualization} or with Docker containers \cite{boakye2023securing}. As a result, these simulations can typically run on a single computer, offering researchers a flexible and easily deployable testbed. In addition, cyber-attacks that would typically cause environmental damage to a real system can be safely investigated.

Dehlaghi-Ghadim et al. \cite{dehlaghi2023icssim} developed a framework for building virtual ICS simulations based on an Object-Oriented Programming (OOP) approach using the Python programming language. Their system utilities Docker containerization technology to replicate ICS components in a deployable environment. However, a prominent flaw within their framework is its applicability for creating custom ICS simulations. Their OOP approach makes it difficult to create custom simulations, as low-level coding and class adjusting is required to do so.

Boakye-Boateng et al. \cite{boakye2023securing} follow a similar approach as they also utilize Docker containers and Python scripts to construct an electrical substation testbed. Their use of Python for implementing direct logical control introduces limitations similar to those identified in the framework developed in \cite{dehlaghi2023icssim}, mainly in terms of reduced flexibility for manual adjustment and independent component implementation.

Another ICS simulation framework has been developed in \cite{alves2018virtualization} that uses virtual machines to realize different ICS components. The framework includes a suite of open-source software tools that can be used to simulate different ICS device functionalities, such as OpenPLC for PLC implementation, ScadaBR for Human Machine Interface (HMI) devices, and Simulink for modeling physical processes. These tools are deployed on virtual machines to construct the simulation environment. A key limitation of virtual machine-based approaches is computational limits. Unlike containers, which share an Operating System (OS) kernel, virtual machines virtualize an entire OS which results in high computational requirements for larger simulations.

The authors of \cite{morris2015industrial} also use virtual machines to construct a virtual gas pipeline testbed. Similarly to the framework in \cite{alves2018virtualization}, Simulink is used to virtualize the physical process, which involves modeling a pump, valve, pipeline and fluid flow.

de Brito and de Sousa Jr \cite{de2022development} present another virtual machine-based testbed that simulates a much more complex control system: a nuclear power plant. They utilize the Asherah Nuclear Power Plant (NPP) simulator to implement the physical system of a condenser cooling pump, which is a specific sub-system of a nuclear power plant. Similarly to previous testbeds, they use OpenPLC for PLC implementation and ScadaBR to construct a HMI component.

TinyICS \cite{candia2024tinyics} is another open-source ICS simulator built on NS-3 that models industrial communication using the Modbus protocol and provides Python bindings for ease of use by control system engineers. While TinyICS focuses on simulating process behavior and capturing communication traffic in customizable industrial scenarios, our ICS-SimLab platform extends beyond single-system emulation to support the rapid deployment of diverse, Docker-based ICS architectures aligned with the Purdue Enterprise Reference Architecture, explicitly targeting cyber security experimentation and intrusion detection system (IDS) development.

The authors of \cite{formby2018lowering} present a Graphical Realism Framework for Industrial Control Systems (GRFICS), which is a framework that utilizes the Unity game engine to visualize ICSs. They simulate the Tennesse Eastman Process (TEP), a widely recognized industrial process originally designed for evaluating process control and monitoring methods \cite{bathelt2015revision}. Their simulation development has been developed primarily as an educational resource where researchers and student can explore difference ICS attacks and defense mechanisms in the visual environment.

\subsection{Hybrid Simulations}
Physical testbeds and virtual testbeds both have their benefits and drawbacks. Many researchers utilize the benefits of both approaches through developing hybrid ICS testbed. Such a testbed utilizes real industrial hardware, such as Schneider PLC models, in conjunction with simulated physical processes to gain high levels of simulation accuracy without having to fully develop the physical control system.

The authors of \cite{mubarak2022industrial} developed a hybrid waste water simulation using industrial components combined with process simulation modules. They used a SIMATIC S7-1200 Siemens PLC, SIMATIC Basic Panel HMI and an Ethernet switch to implement the supervisory layers along with simulation modules with analogue I/O for the physical waste water processes. Many hybrid ICS testbeds follow a similar approach where only the physical levels are simulated using software applications.

A hybrid simulation developed in \cite{noorizadeh2021cyber} simulates the Tennessee Eastman Process, similarly to \cite{formby2018lowering}. In this testbed, the TEP environment is simulated using software on a dedicated PC with I/O modules connected to real PLCs. Their testbed demonstrates how a complex and well-established industrial process can be simulated in a hybrid setup while incorporating actual industrial hardware.

CPS-DERMS \cite{kondu2025cps} introduced a sophisticated multi-co-simulation hardware-in-the-loop testbed that integrates OPAL-RT for grid-side simulations, NS-3 for large-scale cyber simulations, and a custom Python-based DERMS platform to study vulnerabilities in DER-DERMS communication. While their focus was on demonstrating novel Man-in-the-Middle (MITM) attack strategies within power distribution systems, including payload manipulation and stealth techniques, our work with ICS-SimLab is oriented toward simulating cyber-physical threats in industrial control systems (ICS) within industrial plant environments.

\subsection{Limitations and Areas for Improvement}
Most ICS testbeds in literature have been designed to simulate one specific type of control system. There has been little work directed towards developing a generalized system capable of simulating a broad range of control systems architectures.

Furthermore, the current landscape of ICS testbeds offer researchers limited flexibility when it comes to developing custom components and extensions. In physical and hybrid setups, this limitation is understandable as hardware must be purchased and configured to match the testbed environment. However for virtual simulations, having a system that allows researchers to configure and customize their own testbeds would be of great value to the research community. It would allow for the development of cyber security solutions across diverse control system architectures, leading to more accurate and generalizable solutions.

The current virtualization technologies can also be improved to increase the efficiency of simulation deployment. Most virtual testbeds utilize virtual machines to implement different ICS devices. Whilst virtual machines offer accurate encapsulation, running many of them at the same time has significant computational cost. Containers are useful alternatives to virtual machines as they instead virtualize at the OS level, meaning that they share kernel resources and don’t virtualize the OS itself. They provide the same level of abstraction that virtual machines provide, but are far more lightweight and portable.

\section{ICS Architecture Overview}
\label{sec:background}
Industrial Control Systems consist of various different components. These components have unique purposes and roles which vary between systems. A typical ICS setup involves a network of these components communicating amongst each other to perform overall process control. As explored previously, the variety of ICS environments across industrial applications differ significantly. Despite these differences, there are several shared characteristics amongst all control systems. These characteristics typically include common industrial devices, communication protocols, and topology architectures such as the Purdue Enterprise Reference Architecture \cite{williams1994purdue}.

\subsection{Components}
There exists many ICS devices that are used amongst all control systems for similar operations. Common ICS components and modules are described in the following sections. These components can all be simulated within ICS-SimLab.

\subsubsection{Programmable Logic Controllers (PLC)}
PLCs are responsible for handling logic and switching operations \cite{drias2015analysis}. They manage field devices such as sensors and actuators, and perform tasks related to process control and data acquisition. A PLC operates by receiving input signals, executing predefined logic based on those signals, and then transmitting output signals to other devices. The IEC 61131-3 standard defines the programming languages used for PLC development \cite{alphonsus2016review}, with Ladder Logic being the most common one.

\subsubsection{Human Machine Interfaces (HMI)}
HMIs act as centralized control points within an ICS \cite{igure2006security}. They act as clients to PLCs, and are used to request data for visualization and send commands for manual control.

\subsubsection{Sensors and Actuators}
These components are field devices that monitor and handle physical interactions. Sensors are devices that monitor process measurements, such as temperature and pressure. Actuators are responsible for physical movements in process control, and included devices such as valves and conveyor belts. 

\subsubsection{Hardware-in-the-Loop (HIL) Modules}
Whilst not found in real-world ICS environments, HIL components are used to simulate physical processes through complex mathematical models. In industrial contexts, HIL modules emulate physical interactions to enable testing of operational components such as PLCs. While HIL components are not deployed in real systems, they are used within ICS testing, and can also facilitate physical processes for ICS testbeds.

\subsection{Modbus Protocol}
Modbus is the most widely used communication protocol for Industrial Control Systems \cite{7129084}. The Modbus protocol is an open-source protocol used for communication amongst ICS devices. The simplicity and applicability of Modbus has made it a favored protocol within industry.

Modbus has a master/slave architecture. Modbus-compliant devices can either act as a master or as a slave: master devices send Modbus ``requests'' and slave devices reply with Modbus ``responses'' \cite{dutertre2007formal}. There are three variants of the protocol: RTU, ASCII and TCP. Modbus-RTU and ASCII both operate on serial communication lines. The difference between these variants is that RTU utilizes binary encoding whilst ASCII uses human-readable characters. Modbus-RTU is preferred within industry for serial communication due to higher efficiency and more widespread support. As such this paper only addresses Modbus-RTU for serial communication.

\begin{figure}[htbp]
    \centering
    \includegraphics[width=\linewidth]{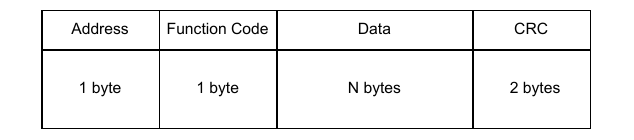}
    \caption{Structure of the Modbus-RTU frame}
    \label{fig:modbus_rtu}
\end{figure}

The Modbus-RTU frame consists of four fields as shown in Figure \ref{fig:modbus_rtu}: Address, function code, data, and a Cyclic Redundancy Check (CRC) field as follows
\begin{enumerate}
    \item Address: One byte used to assign unique addresses to Modbus devices.
    \item Function Code: One byte used to specify a predefined function that dictates the purpose of a Modbus request.
    \item Data: A variable-length field that contains parameters for Modbus requests or return values for Modbus responses.
    \item CRC: Two bytes for CRC error detection.
\end{enumerate}

Modbus-TCP is an encapsulation of the RTU variant into a TCP/IP packet. It facilitates the use of the Modbus protocol over networks operating on the TCP/IP stack \cite{swales1999open}. Essentially, it is the same as the RTU variant but with all the features of the TCP/IP stack. There are some additional fields involved with the TCP variant, but overall the key functionality remains the same. Due to the mainstream popularity of the Modbus protocol, it has been used to facilitate inter-device communication within ICS-SimLab.

\subsection{Purdue Enterprise Reference Architecture}
\begin{figure}[htbp]
    \centering
    \includegraphics[width=\linewidth]{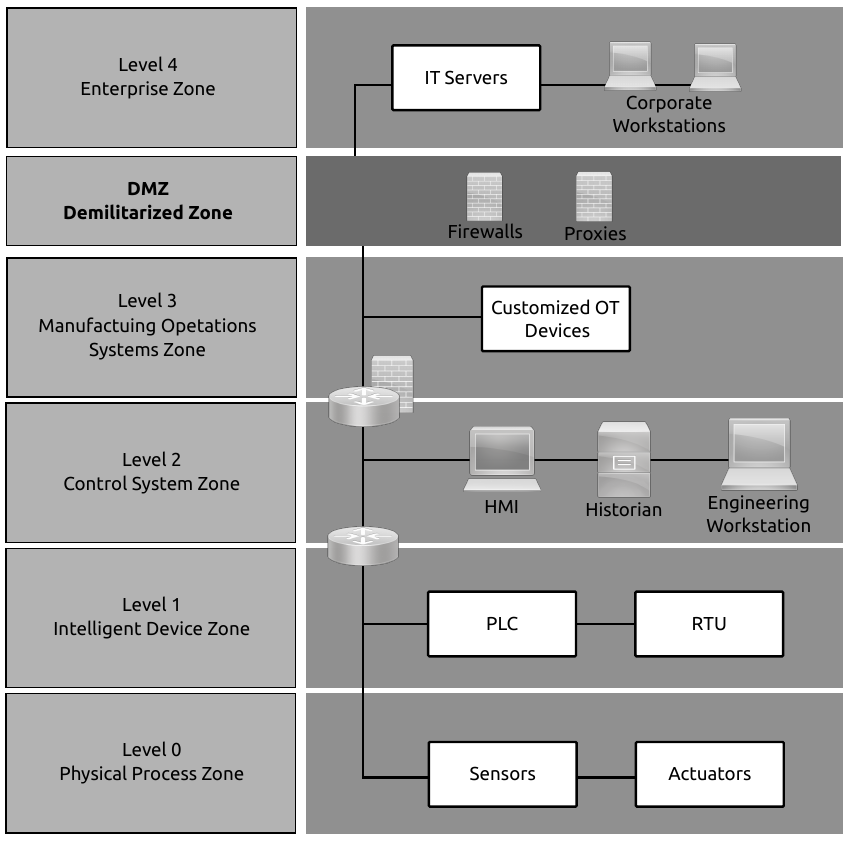}
    \caption{Levels of the Purdue Enterprise Reference Architecture}
    \label{fig:purdue}
\end{figure}

Industrial Control Systems tend to follow an architecture called the Purdue Enterprise Reference Architecture \cite{williams1994purdue}. This model describes an ICS in a layered approach, with each layer representing a different level of component interaction. Lower layers interact with the physical environment of the system, and high layers handle logic and supervision. Figure \ref{fig:purdue} illustrates the levels of this model and highlights where certain ICS components belong in the architecture. ICS-SimLab follows the Purdue model when constructing the architecture of configured ICS simulations.

\section{ICS-SimLab Design and Implementation}
\label{sec:sim-method}
In this section we describe the design and implementation process of the ICS-SimLab simulation software. We elaborate on the reasoning behind key design choices and highlight innovations and features that can improve current ICS simulation practices. Additionally, we present three preconfigured simulations built using ICS-SimLab which can be readily used as immediate ICS testbeds.

\subsection{Device Implementation}
ICS-SimLab is capable of simulating all modules described in Table \ref{tab:components}. These components have been selected as they represent common devices responsible for core functionality in most ICS environments, with the exception of HIL modules which are typically used only in simulation and testing scenarios.

\begin{table}[htbp]
    \caption{Components simulated by ICS-SimLab}
        \begin{center}
                \begin{tabular}{|p{2.6cm}|p{4.5cm}|}
                \hline
                \textbf{Component} & \textbf{Description}  \\
                \hline
                Human~Machine Interfaces (HMI) & Displays collected data and enables manual control. \\
                \hline
                Programmable Logic Controllers (PLC) & Performs control logic for sensors and actuators. \\
                \hline
                Sensors & Reads environmental data from the HIL modules. \\
                \hline
                Actuators & Alters physical processes from the HIL modules. \\
                \hline
                Hardware-in-the-Loop (HIL) Modules & Simulates physical processes. \\
                \hline
            \end{tabular}
        \label{tab:components}
    \end{center}
\end{table}

To achieve device virtualization and isolation, Docker containers have been used. The framework dynamically constructs containers for each component, enabling encapsulation and preventing direct communication between devices. This isolation is crucial for creating a realistic simulation environment, as in practice these components would be physically separated. To assist with running multiple containers in parallel, Docker Compose has been used.

Alternative virtualization methods, such as virtual machines, can introduce drawbacks related to computational cost and deployment complexity \cite{de2022development}. Unlike virtual machines, which virtualize the entire physical computer, containers only virtualize the application environment. They share kernel resources whilst still providing process isolation.

\subsection{Network Emulation}
To ensure accurate ICS communication simulation, Modbus-TCP and Modbus-RTU protocol variants have been implemented into ICS-SimLab. Modbus-TCP utilizes the TCP/IP stack, and so requires a network configuration. Modbus-RTU uses serial port communication.

To virtualize a TCP/IP network, Docker Compose has been used. Compose has in-built functionality that can+ create networks for containers to use. ICS-SimLab uses this feature for Modbus-TCP communication. For Modbus-RTU, the \texttt{socat} utility has been used. \texttt{socat} has the ability to construct virtual serial ports, which ICS-SimLab then uses as a means for Modbus-RTU communication amongst containers. Note that \texttt{socat} has been primarily built for Unix-specific systems, and so will not work on native Windows.

Docker containers by default have network interfaces, which allow them to communicate via IP. However, for serial communication, container volumes have been deployed as a means for parsing virtual serial port input into the containers.

Sensors and actuators require a means of communicating with the physical environment, which in ICS-SimLab is simulated by the HIL modules. Similar to the hardware-aware layer presented in \cite{dehlaghi2023icssim}, ICS-SimLab uses a shared SQLite database among sensors, actuators, and HIL modules. Sensors read from the database, while actuators write to it. The physical values themselves are defined within a JSON configuration file, as described in Section \ref{subsect:config}. The HIL modules modify these physical values based on customized logic implemented in Python.

\subsection{Configuration System}
\label{subsect:config}

ICS-SimLab uses a JSON configuration file to define all components of a simulation. Only one JSON file is needed per ICS type. For example, one file may define the components of a water storage tank system, while another specifies the layout of a gas pipeline simulation. Users can create their own JSON files to specify the components and functionalities they wish to incorporate. ICS-SimLab will then parse the JSON file and deploy a simulation based on the provided configuration. This configuration-based approach allows ICS-SimLab to efficiently deploy a variety of ICS architectures, and more importantly enables researchers to customize their own ICS testbed. 

The JSON file includes parent-level properties that correspond to the different devices defined in Table \ref{tab:components}. Within these properties, further configurations are required to define how the device works. Table \ref{tab:configurations} describes all of the key device-related properties that can be defined.

\begin{table}[htbp]
    \caption{Key property configurations in the JSON file}
        \begin{center}
                \begin{tabular}{|p{2cm}|p{3.4cm}|p{1.9cm}|}
                \hline
                \textbf{Attribute} & \textbf{Description} & \textbf{Components}  \\
                \hline
                \texttt{name} & Custom name for the component. & All components \\
                \hline 
                \texttt{network} & IP and relevant Docker network interface. & All components \\
                \hline
                \texttt{inbound \_connections} & Modbus slaves/server configurations. & PLC, sensors, actuators \\
                \hline
                \texttt{outbound \_connections} & Modbus master/client configurations. & HMI, PLC \\
                \hline
                \texttt{registers} & Defines used Modbus register addresses. & HMI, PLC, sensors, actuators \\
                \hline
                \texttt{monitors} & Defines inter-device reading communication. & HMI, PLC \\
                \hline
                \texttt{controllers} & Defines inter-device writing communication. & HMI, PLC \\
                \hline{}
                \texttt{logic} & Specifies a Python file for custom logic implementation. & PLC, HIL \\
                \hline
                \texttt{physical \_values} & Specifies environmental values. & HIL \\
                \hline
            \end{tabular}
        \label{tab:configurations}
    \end{center}
\end{table}

Python has been used to develop the core configuration software of ICS-SimLab, including the functionality involved with booting up the various containers using the JSON file. The \texttt{pymodbus} Python library has been utilized to implement all Modbus communication. The containers themselves all run a Python service script with different logic depending on what device the container is simulating. For the components that depend on some predefined logic, such as the PLCs, additional Python scripts are created to provide a means of custom logic implementation, which are referenced in the JSON configuration file.

\begin{figure}[htbp]
    \centering
    \includegraphics[width=\linewidth]{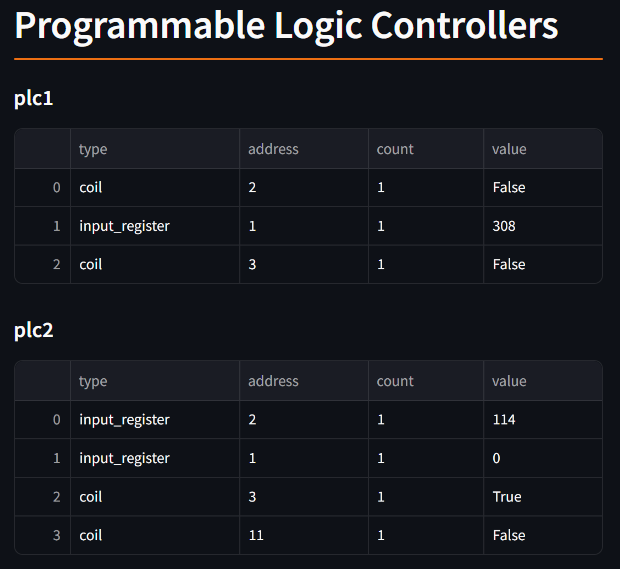}
    \caption{Section of the Streamlit dashboard displaying the addresses and values of PLC registers}
    \label{fig:dashboard}
\end{figure}

Furthermore, the Streamlit web framework has been used to implement a dashboard for visualizing the entire simulation. This dashboard serves a purely observational purpose and is not considered an operational ICS component. Figure~\ref{fig:dashboard} shows an example section of this dashboard interface.

\subsection{Preconfigured Simulations}
\label{subsect:preconfigured}
\begin{figure}[htbp]
    \centering
    \includegraphics[width=0.75\linewidth]{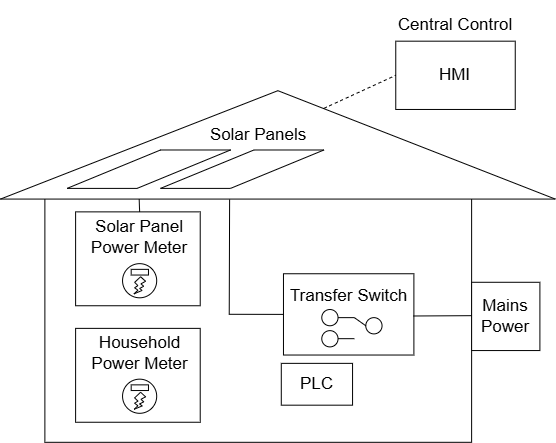}
    \caption{Solar power setup simulation}
    \label{fig:smart_grid}
\end{figure}

To demonstrate the effectiveness of ICS-SimLab in generating ICS simulations, three simulations have been preconfigured. The first simulation represents a solar panel system with an automatic transfer switch. It includes a single HIL module that emulates solar power generation, along with a transfer switch that can alternate the input power between mains and solar. The setup consists of two power meter sensors: one for recording solar power output and one for monitoring the total power input. The transfer switch is the only actuator in this simulation. A PLC has been configured to send control signals to the transfer switch based on the solar power generation level. If the power falls below a defined threshold, the PLC switches to mains power and vice versa. Finally, a HMI device has been implemented for data acquisition across the components. Figure \ref{fig:smart_grid} displays this simulation setup.

\begin{figure}[htbp]
    \centering
    \includegraphics[width=0.75\linewidth]{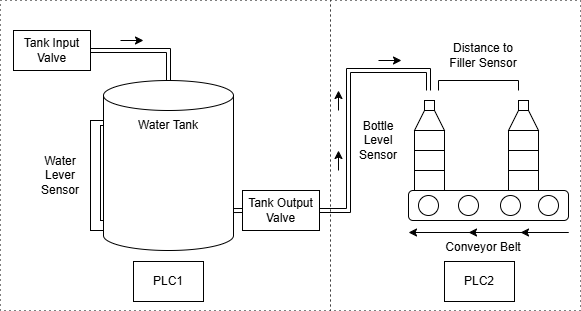}
    \caption{Water bottle filling facility, inspired by the configuration in \cite{dehlaghi2023icssim}.}
    \label{fig:water_bottle}
\end{figure}

The second simulation models a water bottle filling facility, shown in Figure \ref{fig:water_bottle}, and has been replicated from the testbed presented in \cite{dehlaghi2023icssim}. This simulation is divided into two sections, each controlled by a separate PLC. The first PLC manages a water tank with an input valve, output valve, and a water level sensor. The PLC ensures that the tank maintains an adequate water level. The output valve leads to a conveyor belt system that moves empty water bottles underneath it. The second PLC controls this conveyor belt and sends command signals to the first PLC when a bottle is in position and ready to be filled. This simulation has been replicated in ICS-SimLab to demonstrate the platform's flexibility and compatibility with different ICS architectures.

\begin{figure}[htbp]
    \centering
    \includegraphics[width=0.75\linewidth]{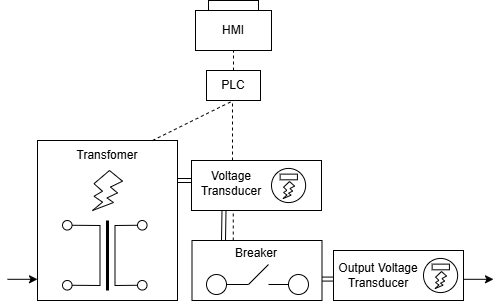}
    \caption{Intelligent Electronic Device setup, inspired by the testbed in \cite{boakye2023securing}.}
    \label{fig:ied}
\end{figure}

Figure \ref{fig:ied} illustrates the third simulation, which models an Intelligent Electronic Device (IED) setup. This simulation has taken inspiration from the testbed presented in \cite{boakye2023securing}, which involves the virtualization of an electrical substation. In this testbed, a simulated transformer undergoes tap changes to alter its output voltage at random intervals, either autonomously or via a HMI device. If the output voltage falls outside an acceptable range, the breaker trips until the voltage returns to a stable value. All of this control logic has been implemented through a PLC.

The source code for ICS-SimLab is publicly available in an online GitHub repository, along with the preconfigured simulation environments. \footnote{\url{https://github.com/JaxsonBrownie/ICS-SimLab}}

\section{ICS-SimLab Cyber Security Applications}
\label{sec:attack-method}
This section outlines the applications of ICS-SimLab in the context of cyber security research. Various ICS-specific cyber attacks are executed on the three preconfigured simulations, which are defined in Section \ref{subsect:config}, to investigate the effects on these attacks. A network data collection system is introduced to capture both benign and malicious activity. Finally, a data labeling system is used to generate datasets that can support the development of Intrusion Detection Systems (IDS) using Machine Learning (ML) techniques.

\subsection{Cyber-Attacks and Vulnerabilities}
ICS environments are vulnerable to attacks unique to their architecture, especially when compared to other systems such as traditional IT systems \cite{drias2015analysis, stouffer2011guide}. For instance, the Modbus protocol used for ICS device communication lacks security controls common to modern IT environments, such as encryption and authorization mechanisms. The lack of encryption means that an attacker can listen in on communications and perform Man-in-the-Middle attacks \cite{7129084}, and since there are no authorization checks attackers can easily send malicious requests given an initial entry point into the ICS system.

\begin{table}[htbp]
    \caption{Implemented cyber attacks based on \cite{morris2013industrial}}
        \begin{center}
                \begin{tabular}{|p{2cm}|p{1.8cm}|p{3.5cm}|}
                \hline
                \textbf{Attack Name} & \textbf{Classification} & \textbf{Description}  \\
                \hline
                Address Scan & Reconnaissance & An address scan to determine Modbus-compliant device addresses. \\
                \hline
                Function Code Scan & Reconnaissance & Scans for valid Modbus function codes. \\
                \hline
                Device Identification & Reconnaissance & Sends a Modbus request with function code 0x2B, which can return device-specific information. \\
                \hline
                Naive Sensor Read & Response and Measurement Injection & Scans over all Modbus registers and relays and attempts to find used addresses. \\
                \hline
                Sporadic Sensor Measurement Injection & Response and Measurement Injection & Writes random values to coils/holding registers \\
                \hline
                Force Listen Mode & \raggedright Command Injection & Sends a Modbus request with function code 0x08 and sub-function code 0x0004 to force a device into Force Listen Mode \\
                \hline
                \raggedright Restart Communications & \raggedright Command Injection & Sends a Modbus request with function code 0x08 and sub-function code 0x0001 to restart the device. \\
                \hline
                Data Flood Attack & \raggedright Denial of Service & Floods the network with Modbus packets \\
                \hline
                Connection Flood Attack & \raggedright Denial of Service & Floods the network with TCP connection requests. \\
                \hline
            \end{tabular}
        \label{tab:attacks}
    \end{center}
\end{table}

Morris \cite{morris2013industrial} presents a number of ICS cyber attacks. Table \ref{tab:attacks} displays all the attacks selected from \cite{morris2013industrial} that have been used on the simulations generated with ICS-SimLab. These attacks are specific to ICS environments using the Modbus protocol, and are general enough to be able to be used on any control system type. Whilst it's common for cyber security researchers to develop cyber attacks tightly coupled to specific testbed architecture \cite{dehlaghi2023icssim, de2022development, morris2011control}, we chose to execute attacks that can be used across multiple control systems.

\begin{figure}[htbp]
    \centering
    \includegraphics[width=\linewidth]{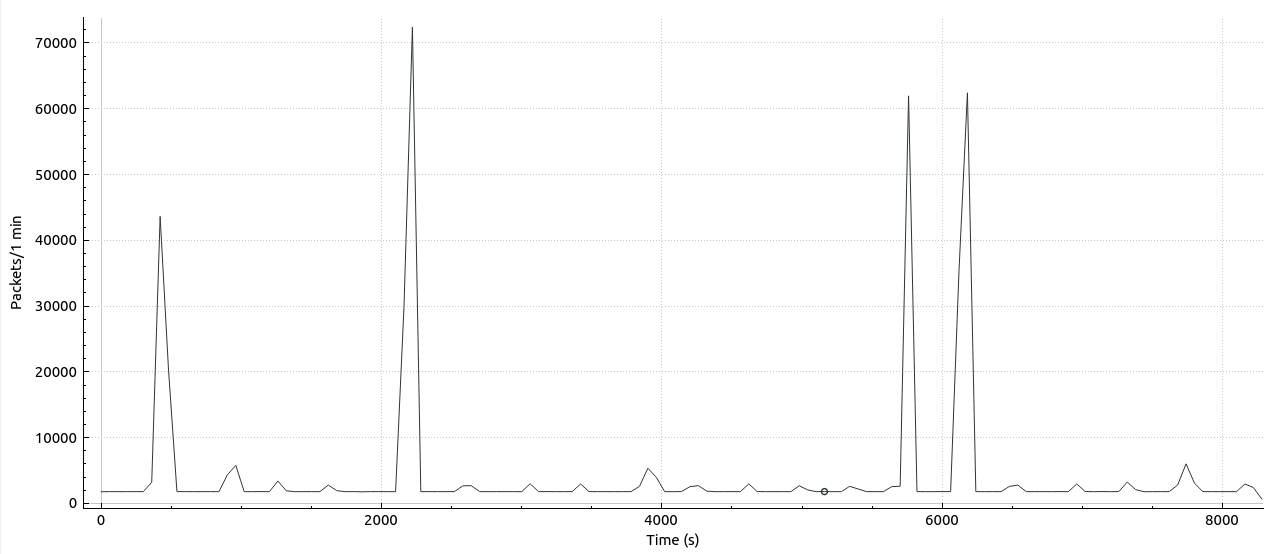}
    \caption{Packets during a connection flood attack.}
    \label{fig:dos}
\end{figure}

These attacks have a clear and expected effect on each of the simulations. As shown in Table \ref{tab:attacks}, reconnaissance attacks have the purpose of gathering device and network information. Response and measurement injection attacks involve crafting malicious Modbus packets for data manipulation. Command injection attacks abuse unsafe standard Modbus function codes to issue unauthorized commands. Finally, denial of service attacks aim to overwhelm the system with excessive network packets to mask genuine network activity. Figure \ref{fig:dos} demonstrates the influx of packets involved in a connection flood attack, as indicated by the peaks in traffic activity.

\subsection{Dataset Generation}
An attacking module was created with Python that would run random attacks from those listed in Table \ref{tab:attacks}. This module was deployed across all of the predefined ICS simulations, with Wireshark being used to capture the resulting network activity. After 30 minutes of simulating both benign and malicious network activity, the captured traffic data was exported to PCAP files for further processing.

\begin{table}[htbp]
    \caption{Features of the generated dataset.}
        \begin{center}
                \begin{tabular}{|p{2cm}|p{4.5cm}|}
                \hline
                \textbf{Feature} & \textbf{Description}  \\
                \hline
                time & Time of packet creation \\
                \hline
                src\_mac & Source MAC address \\
                \hline
                dest\_mac & Destination MAC address \\
                \hline
                src\_ip & Source IP address \\
                \hline
                dest\_ip & Destination IP address \\
                \hline
                protocol & The protocol of the packet. \\
                \hline
                length & Length of the Modbus Application Data Unit (ADU) \\
                \hline
                unit\_id & Modbus unit ID \\
                \hline
                func\_code & Modbus function code \\
                \hline
                data & The Modbus data field \\
                \hline
                attack\_specific & The specific attack used (for malicious packets) \\
                \hline
                attack\_category & Category of the attack (for malicious packets) \\
                \hline
                attack\_binary & Defines if a packet is benign (0) or malicious (1) \\
                \hline
            \end{tabular}
        \label{tab:features}
    \end{center}
\end{table}

A custom data processing script was used to extract relevant features from the PCAP files and construct a CSV-formatted dataset. Each packet was labeled as either malicious or benign, along with its corresponding attack category. Table \ref{tab:features} describes all the features extracted for the dataset. Certain protocols have also been removed from the PCAP file that were unrelated to the main ICS device activity. The PCAP files and generated CSV files can be found in the GitHub repository for the source code for ICS-SimLab, along with the custom data processing script.

Since ICS-SimLab can simulate various ICS environments, it enables the generation of multiple datasets representing different architectures under attack. These datasets can be used collectively to develop a generalized IDS capable of detecting malicious activity across various systems. At the least, having access to multiple ICS environments allows researchers to evaluate the robustness of certain ML techniques against varying network behavior. Additionally, training across multiple datasets introduces the effect of data overfitting, which while a challenging issue to face, can ultimately support the development of more resilient IDS models by exposing their generalization limits.

\section{Conclusion and Future Works}
\label{sec:conclusion}
This paper has presented ICS-SimLab, a software framework for generating ICS simulations through customizable configurations. Containerization technology has been used to enable entire simulations to run efficiently on a single host machine. Furthermore, a JSON-based configuration system allows researchers to construct custom ICS testbed environments, which can facilitate more effective cyber security research by enabling the development and evaluation of solutions across diverse ICS architectures.

To demonstrate the capabilities of ICS-SimLab, three preconfigured simulations have been developed: a solar panel smart grid setup, a water bottle filling facility, and an Intelligent Electronic Device (IED) configuration. In addition, ICS-specific cyber attacks were designed and executed on all three simulations, and their effects were observed. Datasets were generated to represent network activity during both normal operations and attack scenarios. These datasets can support the development of network-based Intrusion Detection Systems using techniques such as Machine Learning

For future work, we aim to develop a real-time IDS software capable of detecting ICS-specific attacks. ICS-SimLab can facilitate this development through providing various ICS simulation environments for constructing and validating the IDS system. The datasets generated in this work can be used to train Machine Learning models used for building the IDS. Furthermore, Deep Learning and Generative AI techniques can be investigated to develop more explainable and interpretable IDS models. We intend to thoroughly explore the uses of modern AI approaches to advance beyond traditional machine learning techniques in ICS security.

To enhance the ICS-SimLab framework, future work will explore the use of more accurate tools for virtualizing specific components. Currently, Python is used to simulate physical processes. However, more specialized software such as Matlab or Simulink may offer improved accuracy \cite{alves2018virtualization}. Investigating alternative virtualization approaches may lead to more accurate and realistic simulation results. In addition, since the core functionality of the ICS components, such as the PLCs, is implemented using Python libraries, device-specific attacks cannot be accurately modeled in the current approach. Instead, the containerized approach primarily targets network-level attacks, such as those exploiting Modbus vulnerabilities. Device-specific attacks, such as buffer overflow attacks and firmware-specific exploits, are currently out of scope. In future work, we aim to explore methods for simulating such device-specific attacks while maintaining the benefits of containerized deployment.

\section*{Acknowledgment}
This work is supported by a cross-campus cyber security seed grant from Curtin University.

\balance
\bibliographystyle{ieeetr}
\bibliography{references}

\end{document}